\documentclass[aps,pra,twocolumn,notitlepage,groupedaddress,showpacs,longbibliography,10pt]{revtex4-1} 
\usepackage{amsmath}    
\usepackage{amsfonts}
\usepackage{bm}
\usepackage{amssymb}
\usepackage{appendix}
\usepackage{graphicx}   
\usepackage{color}      
\usepackage{subfigure}  
\usepackage{comment}
\usepackage{siunitx}

\usepackage{soul}
\usepackage[normalem]{ulem}

\newcommand{\im}{\textrm{Im}}

\newcommand{\lin}{l_{\textrm{intra}}}
\newcommand{\lout}{l_{\textrm{inter}}}
\newcommand{\tin}{t_{\textrm{intra}}}
\newcommand{\tout}{t_{\textrm{inter}}}

\mathchardef\mhyphen="2D

\begin{document}

\title{Observation of a higher-order topological bound state in the continuum}
\author{Alexander Cerjan}
\email[]{awc19@psu.edu}
\author{Marius J{\" u}rgensen}
\author{Wladimir A. Benalcazar}
\author{Sebabrata Mukherjee}
\author{Mikael C. Rechtsman}
\email[]{mcrworld@psu.edu}
\affiliation{Department of Physics, The Pennsylvania State University, University Park, Pennsylvania 16802, USA}

\date{\today}

\begin{abstract}
  Higher-order topological insulators are a recently discovered class of materials that can possess zero-dimensional localized states regardless of the dimension of the lattice.
  Here, we experimentally demonstrate that the topological corner-localized
  modes of higher-order topological insulators can be symmetry protected bound states in the continuum;
  these states do not hybridize with the surrounding bulk states of the lattice even in the absence of a bulk bandgap.
  As such, this class of structures has potential applications in confining and controlling light in systems that do not support a complete photonic bandgap.
\end{abstract}

\maketitle

Topological materials have garnered significant interest for their ability to support boundary-localized
states that manifest exotic phenomena, such as the backscatter-free chiral edge states found in Quantum Hall systems \cite{klitzing_new_1980,halperin_quantized_1982,thouless_quantized_1982,buttiker_absence_1988,haldane_model_1988,haldane_possible_2008,wang_observation_2009,umucalilar_artificial_2011,hafezi_robust_2011,fang_realizing_2012,kitagawa_observation_2012,rechtsman_photonic_2013,khanikaev_photonic_2013,hafezi_imaging_2013},
and edge-localized states found in systems with quantized dipole moments \cite{SSH1979, king1993theory, zak1989berry}.
Recently, it was discovered that crystalline symmetries can give rise to a new class of materials
with topological phases that can protect zero-dimensional corner-localized states
in two dimensions, or more generally $d-n$ dimensional states at the boundaries of $d$ dimensional lattices, with $n \ge 2$
\cite{benalcazar_classification_2014,benalcazar2017quad,benalcazar2017quadPRB, song_densuremath-2-dimensional_2017,langbehn_reflection-symmetric_2017,schindler_higher-order_2018,oded2018,wieder2018,miertcorners,EzawaWannier19,benalcazar2019fillinganomaly,lee2019higher,sheng2019two,schindler2019,petrides_higher-order_2020}.
Coined higher-order topological insulators, if
these systems are also chiral or particle-hole symmetric,
their corner-localized states appear at the center of their energy spectrum.
However, the crystalline symmetries that protect these phases do not necessitate the formation of a bulk bandgap in the middle of the spectrum.
This raises an intriguing question,
namely, do these states remain long-lived in-band resonances, exponentially localized to the corner when they are degenerate with the surrounding bulk bands,
or do they instead hybridize with the bulk states and lose their spatial localization?
If the state remains fully spatially localized to the corner forever, it is a bound state in the continuum (BIC) \cite{original_1929,hsu_bound_2016,friedrich_interfering_1985,plotnik_experimental_2011,weimann_compact_2013,hsu_observation_2013,zhou_perfect_2016,cerjan_bound_2019}, 
where the continuum is formed by the surrounding bulk bands whose states extend throughout the infinite lattice; otherwise
the state becomes a standard resonance, where any energy initially added to the corner will eventually radiate away into the infinite lattice with finite lifetime.
Recently, it was theoretically predicted that a higher-order topological phase can be
used to protect a corner-localized BIC if the system
satisfies additional symmetry requirements \cite{benalcazar_hoti_bic_arxiv}.

Finding a protected zero-dimensional BIC in a material whose topology is only dependent upon the crystalline symmetries of
the lattice presents a significant opportunity in the context of two- and three-dimensional photonic crystals \cite{joannopoulos}. Such states
could be used to realize cavities in low-index photonic crystals, where there are no known crystal geometries which yield complete bandgaps \cite{men_robust_2014,cerjan_complete_2017}.
However, although higher-order topological phases have now been demonstrated in a wide range of different
physical platforms, including microwaves \cite{peterson2018}, photonics \cite{noh2018,mittal_photonic_2019,li_higher-order_2020},
acoustics \cite{serra2018observation,ni_observation_2019,xue_acoustic_2019,xue_realization_2019,ni_demonstration_2019,xue_observation_2020},
electric circuits \cite{imhof_topolectrical-circuit_2018,bao_topoelectrical_2019}, and atomic systems \cite{kempkes_robust_2019},
all of these previous studies have been limited to insulator-like systems, and exhibit
their corner-localized states spectrally isolated from their surrounding bulk bands.

\begin{figure*}[t!]
  \centering
  \includegraphics[width=0.9\linewidth]{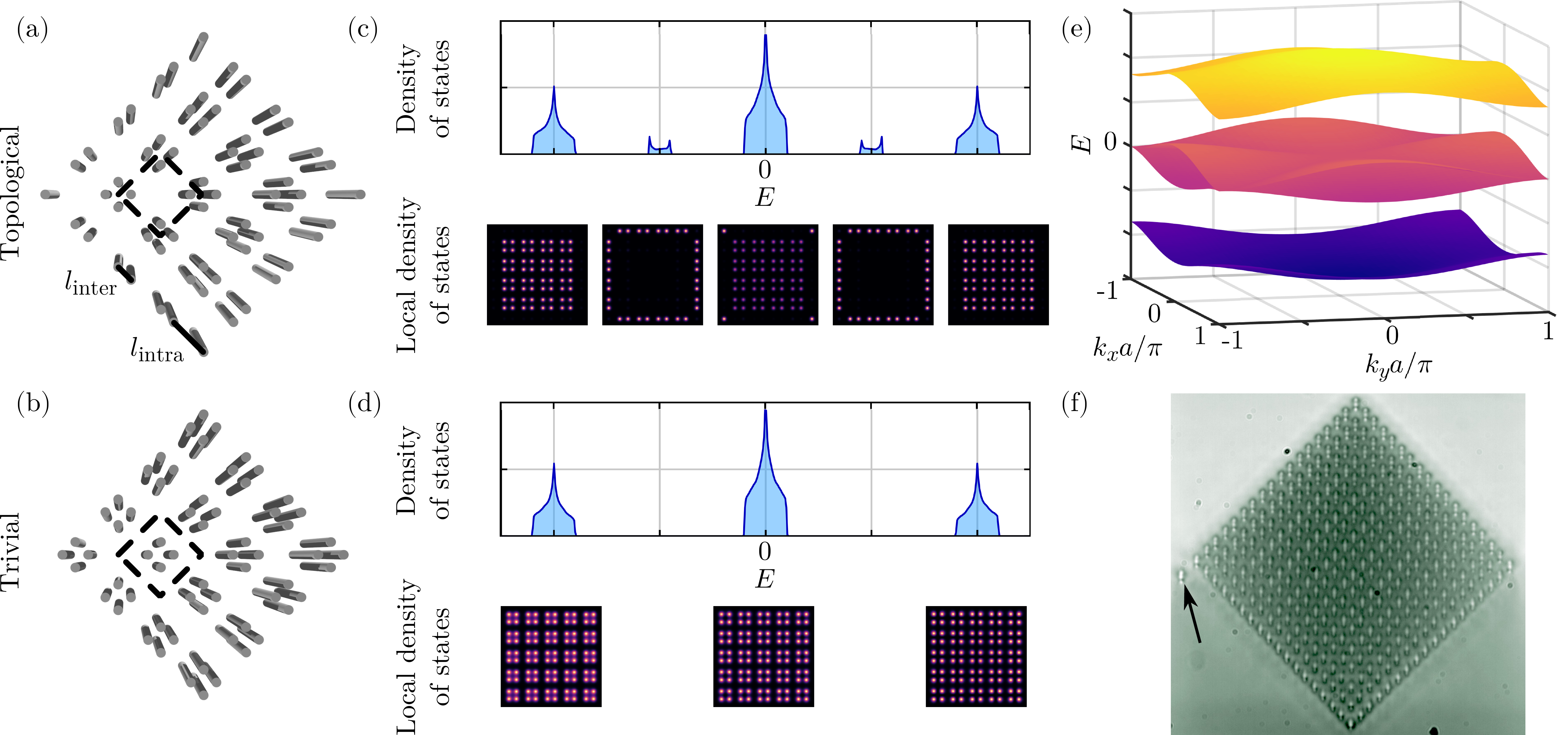}
  \caption{Higher-order topological insulator in a waveguide array.
    (a),(b) Schematic of a higher-order topological insulator in its topological (a) or trivial (b) phase. The unit
    cell of each phase is indicated in the dashed-black square. The distances between adjacent waveguides within a unit
    cell, and between neighboring unit cells, are shown as $\lin$ and $\lout$, respectively. Schematics are not to scale.
    (c),(d) Density of states (top panel) and associated local density of states (bottom panel) for each band for the topological and trivial phases
    of the higher-order topological insulator in finite geometries, respectively. The density of states is calculated using the tight-binding approximation.
    (e) Bulk band structure for the higher-order topological insulator, which is identical for both the topological and trivial phases, calculated using
    full-wave numerical simulations for $\lambda = \SI{850}{\nano\meter}$, $\lin = \SI{17}{\micro\meter}$, and $\lout = \SI{13}{\micro\meter}$.
    (f) White light transmission micrograph of the output facet of a waveguide array with $\lin = \SI{13}{\micro\meter}$ and $\lout = \SI{11}{\micro\meter}$.
    An auxiliary waveguide into which light can be injected, $\SI{20}{\micro\meter}$ away from the array, is indicated with a black arrow.
  }
  \label{fig:intro}
\end{figure*}

Here, we experimentally realize a higher-order topological bound state in the continuum using a two-dimensional waveguide array comprised
of evanescently-coupled waveguides \cite{davis_writing_1996,szameit_discrete_2010}. To show that our waveguide array possesses a BIC, we perform three separate
experiments. First, by injecting light into the corner of the array, we prove that the lattice exhibits a corner-bound mode when the lattice is in its topological phase,
and that this mode disappears across the topological phase transition. Second, by using an auxiliary waveguide to couple into the array, which fixes the effective
energy of the initial excitation, we show that this corner-localized mode appears at zero energy, and is degenerate with bulk states of the lattice.
For consistency with previous studies, we refer to `zero energy' as the propagation constant / energy of a single waveguide, which for chiral-symmetric lattices is at the center of the spectrum.
Finally, we show that our
bound state transforms into a resonance when we break chiral symmetry by detuning the index of refraction of the members of one sublattice. Together, these experiments
prove that the corner-localized state of our higher-order topological waveguide array is a symmetry-protected BIC, and does not hybridize with the bulk bands
so long as the necessary symmetries remain intact.

Our experimental array consists of a square lattice in which each unit
cell contains four waveguides and is $C_{4v}$ symmetric, as shown in Fig.\ \ref{fig:intro}a,b \cite{liu_novel_2017,benalcazar2019fillinganomaly,benalcazar_hoti_bic_arxiv,chen2019}.
As each waveguide within the lattice only supports
a single bound mode for the wavelengths we consider, and the coupling between waveguides decreases exponentially with increasing separation,
our waveguide array can be approximated using a tight-binding model with only nearest-neighbor couplings, such that the lattice is chiral (sublattice) symmetric.
The diffraction of light through the structure is governed by
\begin{equation}
  i \partial_z |\psi(z,\lambda)\rangle = \hat{H}(\lambda) |\psi(z,\lambda)\rangle.
\end{equation}
Here, $|\psi(z,\lambda)\rangle$ is the envelope of the electric field on each of the waveguides at propagation distance $z$ and wavelength $\lambda$.
The coupling coefficients, $\tin$ and $\tout$,
in $\hat{H}$ are determined by the spacings between neighboring waveguides within the same unit cell, $\lin$, and between
adjacent unit cells, $\lout$. 

For an array which is infinite in the transverse plane, the Bloch Hamiltonian of the lattice can be written as
\begin{eqnarray}
  &h(k_x,k_y) = \left( \begin{array}{cc}
    0 & Q \\
    Q^\dagger & 0 \end{array} \right), \label{eq:h} \\
  &Q = \left( \begin{array}{cc}
    t_{\textrm{intra}} + t_{\textrm{inter}} e^{i k_x a} & t_{\textrm{intra}} + t_{\textrm{inter}} e^{i k_y a} \\
    t_{\textrm{intra}} + t_{\textrm{inter}} e^{-i k_y a} & t_{\textrm{intra}} + t_{\textrm{inter}} e^{-i k_x a} \end{array} \right), \label{eq:Q}
\end{eqnarray}
in which $a$ is the lattice constant. To assist with comparisons with the topological literature, we will refer to the eigenvalues
of the waveguide array, $\hat{H}$, as energies, $E$, while noting that physically these values correspond to shifts in momentum, $\beta = -E = k_z - k_0$, of $| \psi \rangle$ along
the $z$ axis. Here, $k_0 = \omega n_0 / c$, where $n_0$ is the index of refraction of the borosilicate glass into which the waveguides are fabricated and
$\omega$ is the frequency of the injected light. 
As all of these modes are bound modes of the waveguides, `zero energy' refers to the energy at the middle of this spectrum.
  
\begin{figure*}[t!]
  \centering
  \includegraphics[width=0.85\linewidth]{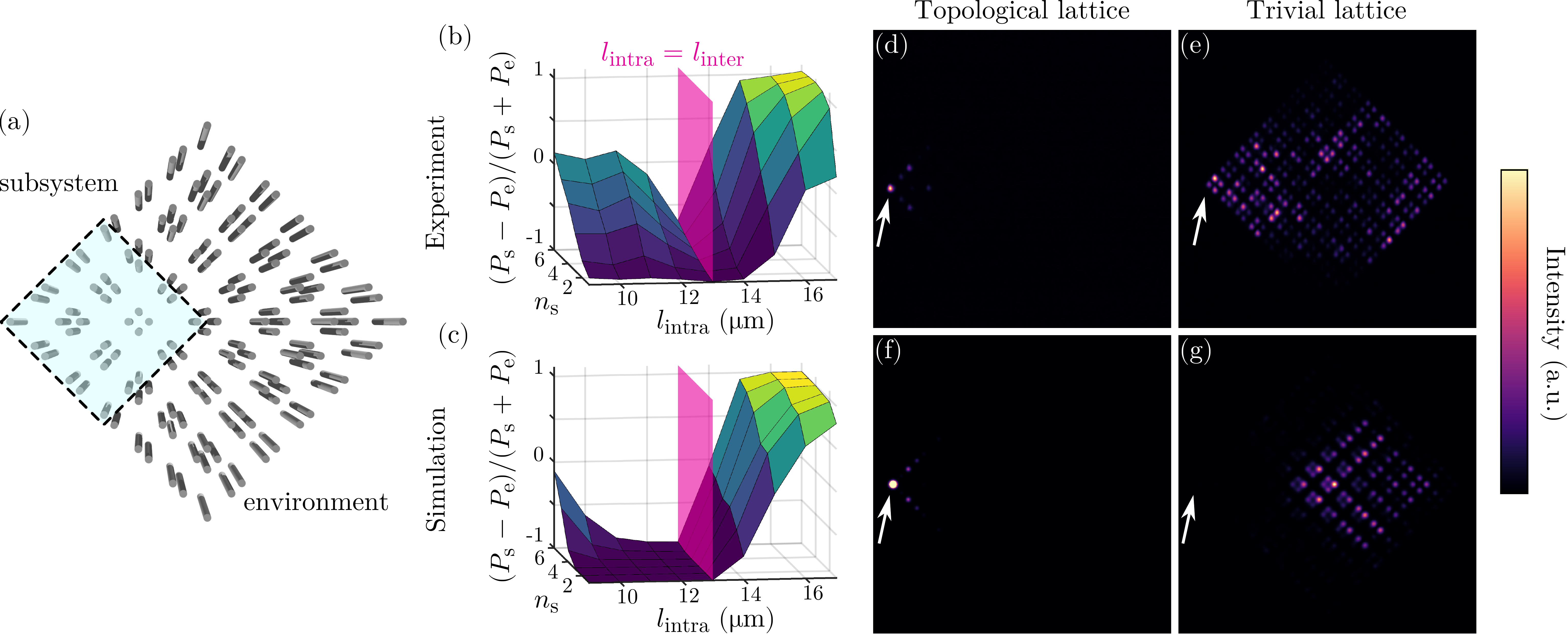}
  \caption{Bound state in a higher-order topological insulator.
    (a) Schematic of a waveguide array in the topological phase with the boundary of the `subsystem' for $n_{\textrm{s}} = 3$ indicated.
    (b),(c) Experimentally observed (b) and numerically simulated (c) fractional power,
    as a function of the size of the subsystem, $n_{\textrm{s}}$, and the spacing between adjacent waveguides within the same unit cell, $\lin$.
    Spacing between adjacent waveguides
    in neighboring unit cells is fixed at $\lout = \SI{13}{\micro\meter}$, and the wavelength of the light is $\lambda = \SI{850}{\nano\meter}$.
    The maximum propagation distance of the array is $L = \SI{7.6}{\centi\meter}$.
    The location of the topological transition is denoted as a magenta plane at $\lin = \lout = \SI{13}{\micro\meter}$.
    (d) Experimentally observed intensity at the output facet for $\lin = \SI{17}{\micro\meter}$. Light is injected into the left-most waveguide at the corner of the array, marked
    with a white arrow. (e) Experimentally observed intensity at the output facet for $\lin = \SI{9}{\micro\meter}$.
    (f)-(g) Same as (d)-(e), except for full wave numerical simulations of the waveguide array.
  }
  \label{fig:corner}
\end{figure*}

The presence of $C_{4v}$ symmetry permits two distinct topological phases depending
on the ratio of the relative spacings between neighboring waveguides within and between adjacent unit cells.
When the lattice is in its topological phase, with $\lin/\lout > 1$, the bands possess different
representations of $C_{4v}$ ($C_{2v}$) at the corresponding high-symmetry points in the Brillouin Zone, $\mathbf{M}$ ($\mathbf{X}$ and $\mathbf{Y}$), than at $\boldsymbol{\Gamma}$.
However, when the lattice is in its trivial phase, with $\lin/\lout < 1$, the bands possess the same symmetry representation at all of the
high-symmetry points. The topological phase transition occurs at $\lin/\lout = 1$, when the bulk bandgap closes at the high-symmetry points, allowing
for the exchange of their representations of these crystalline symmetries.
In a finite lattice, these two phases can be distinguished by their density of states,
as well as the associated local density of states
of each band, shown in Fig.\ \ref{fig:intro}c,d. In its topological phase, this lattice exhibits both edge-localized states in its bulk bandgaps protected by $C_2$ symmetry, as well as a
corner-induced filling anomaly also protected by $C_2$~\cite{benalcazar_hoti_bic_arxiv,benalcazar2019fillinganomaly}. In Fig.\ \ref{fig:intro}c, the presence of these
extra corner-localized states can be observed in the local density of states of the central bulk band of the lattice. Note that when $\lin$ is interchanged with $\lout$, both the topological and trivial
phases of the array have the same bulk band structure consisting of four bands, as displayed in Fig.\ \ref{fig:intro}e.
An example of a facet of a waveguide array is shown in Fig.\ \ref{fig:intro}f.

In the presence of $C_{4v}$ and chiral symmetries, the lattice will always have gapless bulk bands at zero energy, regardless of its topological phase.
These same two symmetries also pin the corner-localized states to zero energy, guaranteeing that the states will always
be degenerate with the bulk bands of the array, while simultaneously protecting the corner-localized modes
from hybridizing with the surrounding bulk states \cite{benalcazar_hoti_bic_arxiv}. This protection comes in two parts. First, two combinations of the four
corner states have incompatible symmetry representations
with those of the surrounding bulk bands at zero energy, and thus cannot hybridize with them.
Then, the two remaining combinations of corner localized states must be both rotationally symmetric partners, with the same energy, and chiral
symmetric partners, with opposite energies, forcing their energies to remain pinned at zero.
This prevents these two corner-localized states from hybridizing with the degenerate bulk states to change their energies or modal profiles, and as such
any hybridization of the corner
states with the surrounding bulk states is simply a change in basis
that does not alter their underlying spatially-localized nature.
This means that all four corner states in our lattice are topologically guaranteed to be zero-dimensional symmetry-protected bound states in the continuum.

In our experiment, it is not possible to completely remove the next-nearest-neighbor couplings which exist between the waveguides in the
array, which means our lattice is not perfectly chiral symmetric. However, the decay length of the corner state due to this slight symmetry breaking ($\sim \SI{25}{\meter}$) 
is significantly longer than the propagation length in our experiments, $L = \SI{7.6}{\centi\meter}$, see Supplementary Information. As such,
our experimental array is effectively chiral symmetric. 

Previous studies of systems supporting BICs have identified that BICs can also be protected by separability, i.e.\ that the
Hamiltonian of the system can be divided into two independent sectors, $\hat{H}(\mathbf{r}) = \hat{H}_x(x) + \hat{H}_y(y)$ \cite{hsu_bound_2016}. However,
despite the fact that Eqs.\ (\ref{eq:h})-(\ref{eq:Q}) are separable, separability
is \textit{not} what protects the higher-order topological BICs we consider here. Analytically, one still observes higher order topological BICs
when additional terms have been added to the lattice's Hamiltonian which obey $C_{4v}$ and chiral symmetries but break separability \cite{benalcazar_hoti_bic_arxiv}, see Supplementary Information.

To experimentally prove that our waveguide array contains a higher-order topological BIC, we first inject light into the corner of the array,
and observe whether most of the light remains confined to this corner or diffracts into the bulk. To assess the localization
of the light at the output facet of the array, we divide the array into two regions, the `subsystem' which represents the square of unit cells with side
length $n_{\textrm{s}}$ closest to the corner, as indicated in Fig.\ \ref{fig:corner}a,
while the remaining waveguides comprise the `environment.' This terminology is chosen for consistency with previous
studies of bound states in the continuum, in which the subsystem (where light is confined) and its surrounding radiative environment
are typically physically distinct regions containing different types of structures.
We then compare the total output power observed in the subsystem, $P_{\textrm{s}}$, with that observed in the environment, $P_{\textrm{e}}$,
using the figure of merit $(P_{\textrm{s}} - P_{\textrm{e}})/(P_{\textrm{s}} + P_{\textrm{e}})$. For this `fractional power,'
values near $+1$ correspond to all of the output power being localized in the subsystem, while values
of $-1$ indicate that all of the output power has diffracted into the environment. In Figs.\ \ref{fig:corner}b,c, we show the fractional power as a function of the
topological phase of the array, $\lin/\lout$, as well as the the size of the subsystem, $n_{\textrm{s}}$. Here, we can clearly see that the light remains localized to the
subsystem, regardless of its size, until the arrays approach the topological phase transition, at $\lin/\lout = 1$. The observed intensity at the output facet is
shown for an example of both the topological and trivial arrays in Figs.\ \ref{fig:corner}d-g. 
Note that the increase seen in the fractional power for
large subsystem sizes for some topologically trivial arrays is due to spurious reflections off of some of the
waveguides at the top and bottom of the array, as
well as back-reflections off of the far side of the array. Nevertheless, it is clear
from Fig.\ \ref{fig:corner}e that these arrays do not possess a bound state.
Thus, these results indicate that the topological waveguide array possesses
a bound state that is connected to the topological phase of the lattice, but does not yet prove that the bound state is degenerate with the surrounding bulk bands.

\begin{figure}[t!]
  \centering
  \includegraphics[width=0.85\linewidth]{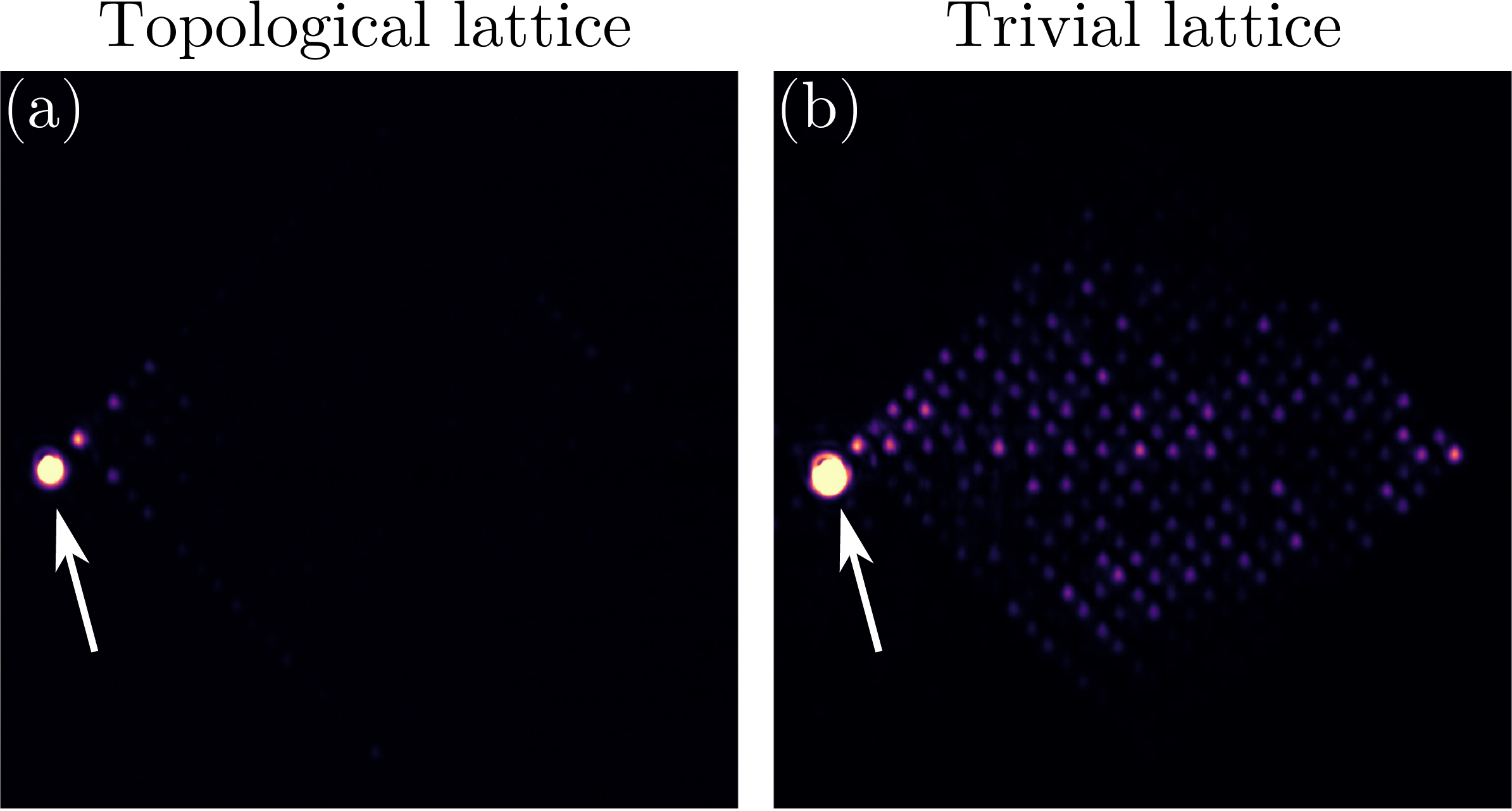}
  \caption{Observation of the BIC and surrounding continuum in a higher-order topological insulator.
    (a) Experimentally observed intensity at the output facet for a topological waveguide array, with $\lin = \SI{13}{\micro\meter}$ and $\lout = \SI{11}{\micro\meter}$.
    Light is injected into the array at $\lambda = \SI{900}{\nano\meter}$ using
    an auxiliary waveguide placed $\SI{20}{\micro\meter}$ away from the corner of the lattice (marked with a white arrow).
    The total length of the array is $L = \SI{7.6}{\centi\meter}$.
    (b) Experimentally observed intensity at the output facet for a trivial waveguide array, with $\lin = \SI{11}{\micro\meter}$ and $\lout = \SI{13}{\micro\meter}$.
  }
  \label{fig:straw}
\end{figure}

\begin{figure*}[t!]
  \centering
  \includegraphics[width=0.95\linewidth]{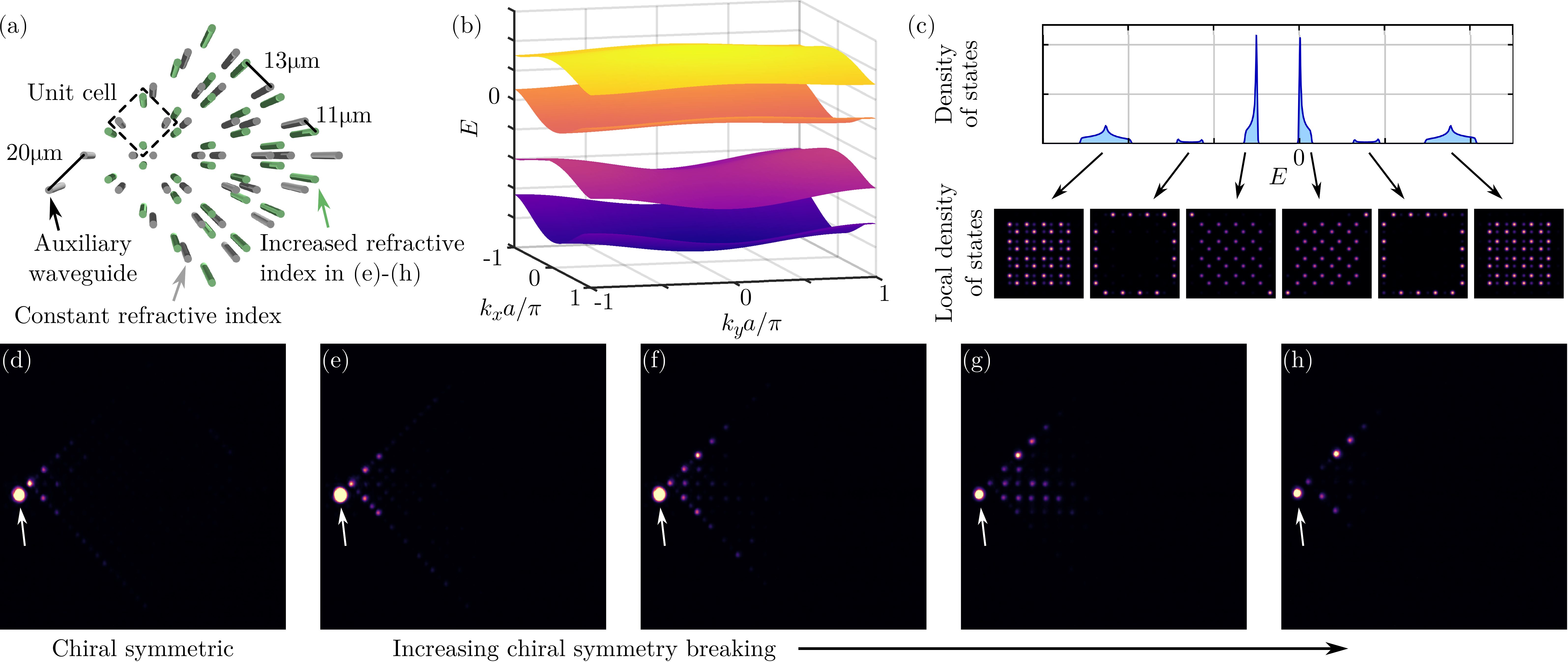}
  \caption{
    Observation of a BIC turning into a resonance as chiral symmetry is broken by detuning the sublattices of the lattice.
    (a) Schematic of a higher-order topological waveguide array with broken chiral symmetry. The auxiliary waveguide where light is
    initially injected is $\SI{20}{\micro\meter}$ away from the waveguide array.
    The waveguides colored in green have been fabricated using slower writing speeds, resulting
    in a larger refractive index, and thus a decreased on-site energy.
    (b) Bulk band structure for the higher-order topological insulator with broken chiral symmetry, calculated using
    full-wave numerical simulations for $\lambda = \SI{900}{\nano\meter}$, $\lin = \SI{13}{\micro\meter}$, $\lout = \SI{11}{\micro\meter}$,
    and in which two of the waveguides have an increased refractive index, $\Delta n_0 = 3.3 \cdot 10^{-3}$, as opposed to $\Delta n_0 = 2.9 \cdot 10^{-3}$.
    These parameters correspond to the experimental results shown in (f), below.
    (c) Density of states (top panel) and associated local density of states for each band (bottom panel) for a tight-binding lattice with broken chiral symmetry.
    Zero energy (center of the spectrum) of the chiral symmetric (unperturbed) array is marked.
    (d) Experimentally observed intensity at the output facet of the symmetric waveguide array, with the same refractive index on all of
    the waveguides, with $\lin = \SI{13}{\micro\meter}$ and $\lout = \SI{11}{\micro\meter}$, for an incident wavelength of $\lambda = \SI{900}{\nano\meter}$.
    The length of the array in the $z$ direction is $L = \SI{7.6}{\centi\meter}$. The auxiliary waveguide where light is injected is marked with a white arrow.
    (e)-(h) Similar to (d), except with increasing the refractive index of the indicated sublattice of the array.
  }
  \label{fig:sublat}
\end{figure*}

To prove that this topological bound state is a BIC, we use a waveguide array with an auxiliary waveguide weakly coupled to the lattice and placed near one of the corners.
Since this waveguide is identical to all others in the lattice, it effectively acts as a fixed zero-energy source.
All light injected into it can only excite states near zero-energy in the array. As can be seen in Fig.\ \ref{fig:straw}a, when the waveguide array is in its topological phase
and the auxiliary waveguide is placed near a corner of the lattice, the dominant excited mode of the waveguide array is the topological corner-localized
mode. However, when $\lin$ and $\lout$ are reversed, the bulk of the lattice remains completely unchanged, but
the array is now in the trivial phase. Upon excitation using an auxiliary waveguide, we see that bulk states of the lattice are excited,
and there is no corner-localized mode. Since the lattice bulk is identical in both cases, we can conclude that there are zero-energy bulk states that
are degenerate with the corner-localized mode in the topological case. This experimentally proves that the corner-localized topological bound states in this array are BICs.

Finally, to demonstrate that this higher-order topological BIC is protected by chiral symmetry, we purposefully break chiral symmetry
by increasing the refractive index on two of the four waveguides in the unit cell, as indicated in Fig.\ \ref{fig:sublat}a, which has the effect of decreasing
the effective on-site energy
of these two lattice sites. This has several effects on the array.
First, this opens a bulk bandgap in the center of the spectrum, in which one of the two central bulk bands remains at `zero-energy' (which is no
longer at the middle of the spectrum), while the other's energy decreases, as shown in Fig.\ \ref{fig:sublat}b. Second, as the modal profile of
each corner-localized state is only supported on two of the four waveguides in the unit cell (diagonally across from one another),
this change also breaks the four-fold degeneracy of the corner-localized states.
Instead, the pair of corner-localized states whose modal profiles overlap with the perturbation decrease their
energy, remaining degenerate with the higher of the two central bulk bands, while the other pair of corner-localized states remain degenerate with the bulk band
at `zero energy' (which is now {\it not} the center of the spectrum). This can be seen in the local density of states for each band of the array, shown in Fig.\ \ref{fig:sublat}c.
However, now that chiral symmetry has been broken, the corner-localized modes are allowed to hybridize with states from their respective bulk bands,
transforming from BICs into resonances of the lattice.
We can observe this transition of one of the BICs into a resonance by incrementally increasing the strength of the sublattice symmetry breaking,
and coupling into the lattice using an auxiliary waveguide, which remains at zero energy, as shown in Figs.\ \ref{fig:sublat}d-h.
As chiral symmetry is lost, as in Figs.\ \ref{fig:sublat}e-h, the wavefunction
within the array begins to disassociate from the corner, and the maximum of this wavefunction travels into the bulk of the array and along the edges,
signifying that all of the states that are being excited by the auxiliary waveguide have significant spatial overlap with the other modes of the lattice.
In other words, the corner-localized state has become a resonance
and is no longer a BIC. This is in clear contrast to what is seen in Fig.\ \ref{fig:sublat}d, where chiral symmetry
is intact and the wavefunction in the lattice remains localized to the corner, indicating the presence of a BIC.

In conclusion, we have experimentally observed a higher-order topological bound state in the continuum in a waveguide array.
This BIC is protected by $C_{4v}$ and chiral symmetries, and is topologically guaranteed to exist at zero energy in the lattice.
Moreover, as these states are able to confine light to a zero-dimensional mode in the absence of a bulk bandgap, which is a greater reduction
in dimensionality than is found in other systems supporting BICs \cite{hsu_bound_2016}, they are a promising candidate
for creating cavities in low-index photonic platforms in the absence of a complete photonic bandgap. Current designs for
photonic crystals that support band gaps require refractive indices of at least $n = 2.1$ in 2D \cite{cerjan_complete_2017}, or $n = 1.9$ in 3D \cite{men_robust_2014}.
In particular, this means that higher order topological BICs could potentially be used to confine light in many important low-index photonic platforms,
such as those based on two-photon lithography in photoresist \cite{freymann_three-dimensional_2010,buckmann_tailored_2012}
and colloids \cite{norris_opaline_2004}, which are typically composed of materials with refractive indices $n \sim 1.5$.  We expect that
zero-dimensional bound states in the continuum of the kind described here will lead to an expanded range of devices in which cavity and defect modes,
for enhancing light-matter coupling, can be found.


\section*{Methods}

We fabricated our waveguide arrays with a Yb-doped fiber laser (Menlo BlueCut) system emitting circularly polarized sub-picosecond ($260$ fs) pulse trains
at $\SI{1030}{\nano\meter}$ with a repetition rate of $\SI{500}{\kilo\hertz}$. The light was focused inside a borosilicate glass (Corning Eagle XG) sample using
an aspheric lens. The borosilicate glass sample was mounted on high-precision $x$-$y$-$z$ translation stages (Aerotech). Each 
individual waveguide was written by translating the glass sample once through the focus of the laser at a speed of $10$ mm$/$s.
This process results in waveguides which only support the fundamental mode, which has an elliptical profile. However, by then orienting the lattice to be
`diagonal' with respect to the surface of the glass slide (i.e.\ the surface of the glass slide lies along the $[1\ 1]$ direction of the lattice), the coupling constants between neighboring
waveguides in both the $[1\ 0]$ and $[0\ 1]$ directions are the same, up to fabrication imperfections. By measuring two-waveguide couplers, we determined that
our waveguides can be modeled in our full-wave numerical simulations using a Gaussian profile,
\begin{equation}
  \Delta n(\mathbf{r}) = \Delta n_0 e^{-r^2/\sigma_r^2}
\end{equation}
with $\Delta n_0 = 2.9 \cdot 10^{-3}$ and $\sigma_r = \SI{4}{\micro\meter}$. For a separation of $\SI{13}{\micro\meter}$ at $\lambda = \SI{850}{\nano\meter}$, this
yields a tight-binding coupling coefficient of $t = \SI{1.62}{\centi\meter}^{-1}$.
For the waveguides used in breaking the chiral symmetry of the lattice as highlighted in Fig.\ \ref{fig:sublat}, the writing speed was incrementally
decreased to be $[8,6,4,2]$ mm$/$s in Fig.\ \ref{fig:sublat}e-h, respectively. These reduced writing speeds can be modeled numerically as shifts of the refractive index
of the waveguides of $\Delta n_0 = [3.1, 3.3, 3.5, 3.7] \cdot 10^{-3}$.

The waveguide arrays were measured using a commercial supercontinuum source (NKT SuperK COMPACT) with a filter to select the desired wavelength (SuperK SELECT). The beam
was focused into the sample using an aspheric lens with an NA of $0.15$ (ThorLabs C280TMD-B) and imaged onto the camera with an achromatic doublet (ThorLabs AC064-015-B-ML).
The images were taken using a CMOS camera (ThorLabs DCC1545M).\\



\begin{acknowledgments}
  This work was supported by the US Office of Naval Research (ONR) Multidisciplinary University Research Initiative (MURI) grant N00014-20-1-2325
  on Robust Photonic Materials with High-Order Topological Protection, the ONR Young Investigator Award under grant number N00014-18-1-2595
  as well as the Packard Foundation under fellowship number 2017-66821.
  W.A.B.\ acknowledges the support of the Eberly Postdoctoral Fellowship at the Pennsylvania State University. M.J. acknowledges the support of
  the Verne M. Willaman Distinguished Graduate Fellowship at the Pennsylvania State University.
\end{acknowledgments}


\pagebreak
\newpage

\onecolumngrid

\section*{Supplemental information for: Observation of a higher-order topological bound state in the continuum}



\setcounter{equation}{0}
\renewcommand{\theequation}{S{\arabic{equation}}}
\setcounter{figure}{0}
\renewcommand{\thefigure}{S{\arabic{figure}}}

\section{Higher order bound state in the continuum without separability}

One known mechanism for creating bound states in the continuum (BICs) is through separability \cite{hsu_bound_2016},
in which the Hamiltonian of the system can be divided into two (or more) parts,
\begin{equation}
  \hat{H}(\mathbf{r}) = \hat{H}_x(x) + \hat{H}_y(y)
\end{equation}
that only depend on a single spatial coordinate. Then, by finding a localized bound state of each individual portion,
for example, $H_x(x) \psi_n(x) = E_n^{(x)} \psi_n (x)$ and $H_y \phi_m(y) = E_m^{(y)} \phi_m(y)$, the combined state,
$\psi_n \phi_m$, is a bound state of $\hat{H}(\mathbf{r})$ with energy $E_n^{(x)} + E_m^{(y)}$, which may reside within
the continuum $\hat{H}(\mathbf{r})$.
Furthermore, one may suspect that this is the origin of the BIC we report in the main text, as the tight-binding model corresponding to our waveguide arrays is
separable.

\begin{figure}[h]
  \centering
  \includegraphics[width=0.95\linewidth]{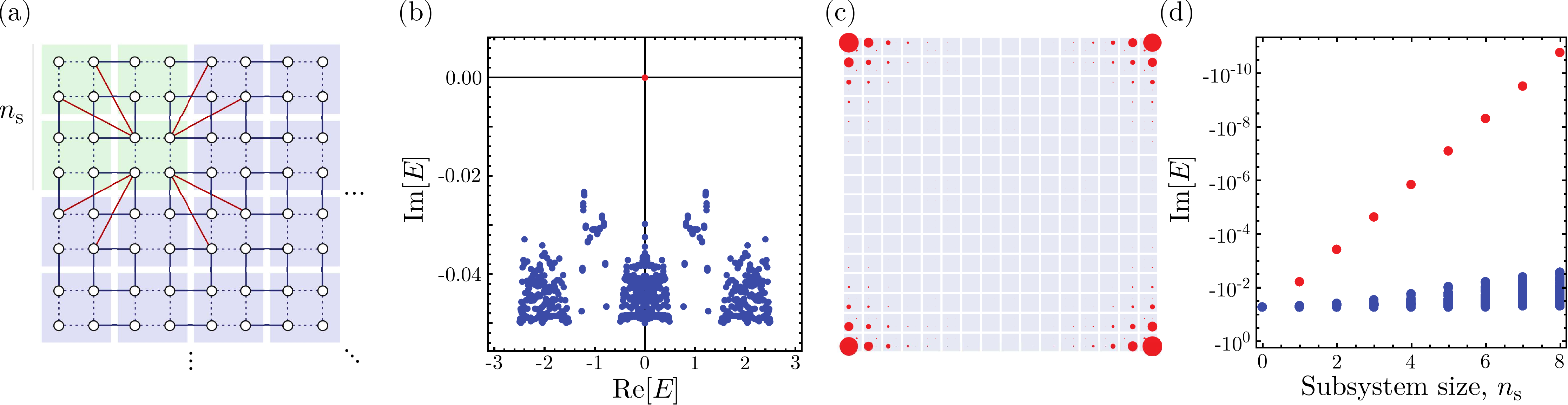}
  \caption{(a) Schematic depicting the tight-binding model used as an example of a non-separable
    higher-order topological BIC. Dashed lines indicate couplings of $\tin = 0.25$, solid black lines indicate couplings
    of $\tout = 1$, and solid red lines indicate chiral-preserving non-separable couplings, $t = 0.025$. Only a single set
    of chiral-preserving non-separable couplings are shown for simplicity. Unit cells are indicated by shaded regions behind
    sets of nodes. Green unit cells are part of the `subsystem,' while blue unit cells are part of the `environment,' and have
    an added on-site loss of $\gamma = 0.05$. Only a single corner of the lattice is shown, but all four corners are
    part of the subsystem for these simulations, and do not have any on-site loss.
    (b) Eigenvalues for this lattice shown in the complex plane for $n_{\textrm{s}} = 3$, i.e.\ the total number of unit cells in each corner which constitute the `subsystem' is $3 \times 3$.
    The lattice has $n_{\textrm{l}} = 16$ unit cells along its sides, for a total lattice size of $16 \times 16$.
    (c) The probability densities of the four eigenstates corresponding to the BICs from (b).
    (d) Plot of the decay rate of the eigenvalues of the lattice as $n_{\textrm{s}}$ is increased.
  }
  \label{fig:sep}
\end{figure}

However, this is not the origin of the protection of the BICs that we observe in our waveguide arrays \cite{benalcazar_hoti_bic_arxiv}. To demonstrate this explicitly,
in Fig.\ \ref{fig:sep} we show numerical tight-binding calculations of a higher-order topological insulator that obeys
both of the necessary protecting symmetries of the BIC, $C_{4v}$ and chiral symmetry, but breaks
separability. This lattice is schematically shown in Fig.\ \ref{fig:sep}a. We then divide this finite lattice into two
regions, the `subsystem,' a region of $n_{\textrm{s}}$ unit cells next to each of the four corners of the lattice, and the
`environment,' which are the remainder of the unit cells in the lattice. A small, but non-zero, amount of loss is then
added to the environment to simulate radiation loss, $0 < \gamma \ll 1$. As can be seen in Fig.\ \ref{fig:sep}b,
this lattice still possesses four essentially real eigenvalues, whose corresponding eigenstates are exponentially
localized to the corners of the lattice, see Fig.\ \ref{fig:sep}c. Finally, by changing the size of the device region,
we can confirm that the energies of these corner-localized states exponentially converge to be real, confirming
that these are BICs, not resonances, as shown in Fig.\ \ref{fig:sep}d.

Thus, even though the waveguide array that we study in the main text is separable, this separability is \textit{not} what protects the
BICs that we observe in our array. Even in the absence of separability, our arrays would still exhibit BICs, so long
as the two necessary protecting symmetries, $C_{4v}$ and chiral, were preserved.

\section{Effect of next-nearest-neighbor couplings in the waveguide array}

As discussed in the main text, in our waveguide arrays it is impossible to completely remove next-nearest-neighbor
couplings between waveguides as the coupling strength is an exponentially decaying function of the spatial separation
between the waveguides, $t(\lambda) = e^{-\alpha(\lambda) l}$, in which $\alpha(\lambda)$ is a wavelength-dependent constant.
In particular, this means that the couplings between diagonally adjacent waveguides in our arrays will be non-vanishing,
and break chiral symmetry, placing a practical limit on the decay length of the corner modes.
However, in practice, the coupling coefficients corresponding to this process are small relative to the dominant energy scale
in the array, i.e.\ the larger of $\tin$ or $\tout$ depending on whether the lattice is in the topological phase. For example,
for the topological lattice shown in Fig.\ 2d,f of the main text, $t_{\textrm{diag}}/\tout \sim 0.08$, where $t_{\textrm{diag}}$
is the next-neighbor coupling strength between waveguides across the diagonal in the unit cell.

\begin{figure}[t!]
  \centering
  \includegraphics[width=0.6\linewidth]{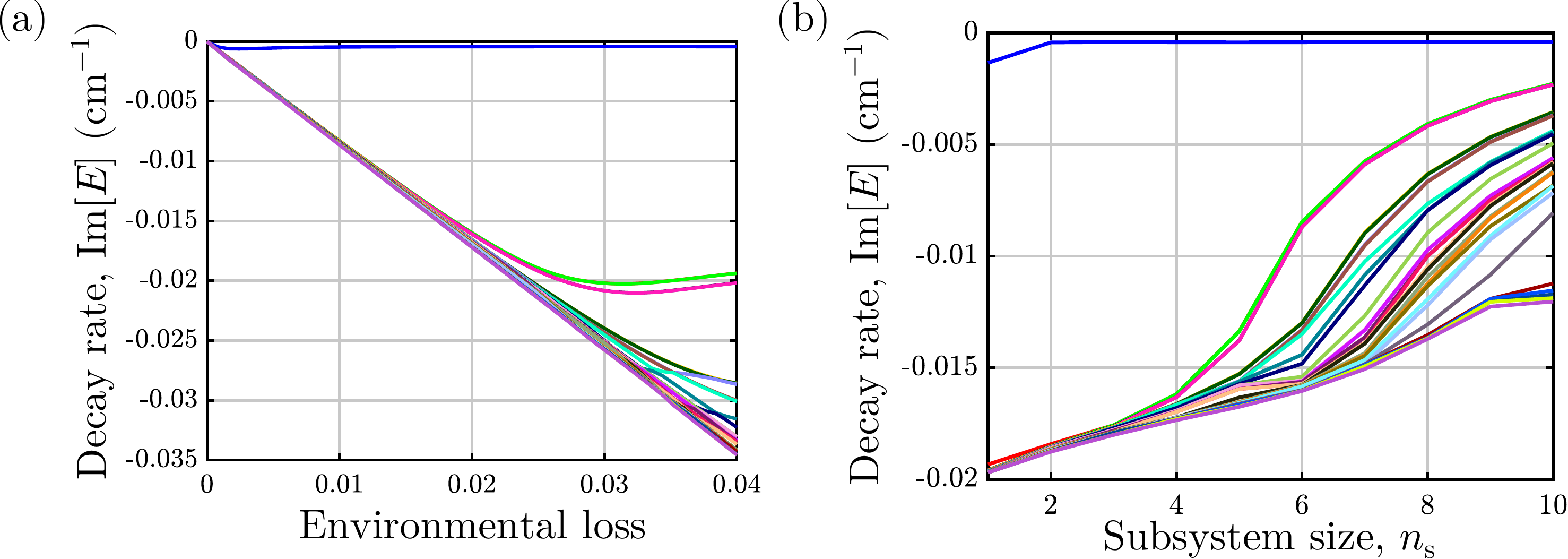}
  \caption{
    (a) Plot of the decay rates of the $30$ least lossy eigenvalues of the lattice as a function of the added loss, $\gamma$,
    to the waveguides in the environment. The plateau seen in the decay rate of the corner-localized
    mode corresponds to its true radiative rate in an infinite lattice. Here, $n_{\textrm{l}} = 32$ (i.e.\ the total lattice size is $32 \times 32$),
    and $n_{\textrm{s}} = 4$. (b) Plot of the decay rates of the $30$ least lossy eigenvalues of the full lattice as a function
    of the edge length of the device region, $n_{\textrm{s}}$ for $\gamma = 0.02$ and $n_{\textrm{l}} = 32$. In both (a) and (b),
    the decay rate of the corner-localized mode is reaching a nearly constant value at approximately $\im[E] = -4 \cdot 10^{-4} \SI{}{\centi\meter}^{-1}$.
  }
  \label{fig:sup2}
\end{figure}

Moreover, we can calculate
the effect of this coupling constant on the now-finite decay length of the corner-localized mode using a tight-binding model with
a small amount of non-Hermitian loss added to an `environment' region. Mathematically, this corresponds to 
\begin{equation}
  \hat{H}_\textrm{tot} = \hat{H} - \sum_{j \in \textrm{env. wgs.}} i \gamma |j\rangle \langle j|,
\end{equation}
in which $\hat{H}$ is given by Eq.\ 1 of the main text, and in which the sum runs over those waveguides in the environment.
As can be seen in Fig.\ \ref{fig:sup2}a, the decay length of the corner-localized
mode quickly saturates as a function of the added loss to the environment, and this plateau corresponds to the radiative loss
of the corner-localized resonance in an infinite lattice without added loss by the Limiting Absorption Principle \cite{v._ignatowsky_reflexion_1905,eidus_LAP_1962,schulenberger_limiting_1971,cerjan_why_2016}.
Moreover, we can confirm this decay length of the corner-localized mode by fixing the added environmental loss
and varying the size of the device region, shown in Fig.\ \ref{fig:sup2}b. In both cases, the decay length converges to approximately $\im[E] = -4 \cdot 10^{-4} \SI{}{\centi\meter}^{-1}$
As stated in the Main Text, for the topological lattice shown in Fig.\ 2d,f, the decay length
is $L_{\textrm{decay}} \sim \SI{25}{\meter}$, which is significantly longer than the waveguide arrays in our experiment ($L=\SI{7.6}{\centi\meter}$). As such,
for the purposes of our experiment, our waveguide array is effectively chiral symmetric.

\twocolumngrid

\end{document}